# Rejoinder: Boosting Algorithms: Regularization, Prediction and Model Fitting

**Peter Bühlmann and Torsten Hothorn**

## 1. DEGREES OF FREEDOM FOR BOOSTING

We are grateful that Hastie points out the connection to degrees of freedom for LARS which leads to another—and often better—definition of degrees of freedom for boosting in generalized linear models.

As Hastie writes and as we said in the paper, our formula for degrees of freedom is only an approximation: the cost of searching, for example, for the best variable in componentwise linear least squares or componentwise smoothing splines, is ignored. Hence, our approximation formula

$$\mathrm{df}(m) = \mathrm{trace}(\mathcal{B}_m)$$

for the degrees of freedom of boosting in the $m$th iteration is underestimating the true degrees of freedom. The latter is defined (for regression with $L_2$-loss) as

$$\mathrm{df}_{\mathrm{true}}(m) = \sum_{i=1}^{n} \mathrm{Cov}(\hat{Y}_i, Y_i)/\sigma_\varepsilon^2, \quad \hat{Y} = \mathcal{B}_m Y,$$

cf. Efron et al. [5].

For fitting linear models, Hastie illustrates nicely that for infinitesimal forward stagewise (iFSLR) and the Lasso, the cost of searching can be easily accounted for in the framework of the LARS algorithm. With $k$ steps in the algorithm, its degrees of freedom are given by

$$\mathrm{df}_{\mathrm{LARS}}(k) = k.$$

For quite a few examples, this coincides with the number of active variables (variables which have been selected) when using $k$ steps in LARS, that is,

$$\mathrm{df}_{\mathrm{LARS}}(k) \approx \mathrm{df}_{\mathrm{actset}}(k)$$
$$= \text{cardinality of active set}.$$

Note that the number of steps in $\mathrm{df}_{\mathrm{LARS}}$ is not meaningful for boosting with componentwise linear least squares while $\mathrm{df}_{\mathrm{actset}}(m)$ for boosting with $m$ iterations can be used (and often seems reasonable; see below). We point out that $\mathrm{df}(m)$ and $\mathrm{df}_{\mathrm{actset}}(m)$ are random (and hence they cannot be degrees of freedom in the classical sense). We will discuss in the following whether they are good estimators for the true (nonrandom) $\mathrm{df}_{\mathrm{true}}(m)$.

When using another base procedure than componentwise linear least squares, for example, componentwise smoothing splines, the notion of $\mathrm{df}_{\mathrm{actset}}(m)$ is inappropriate (the number of selected covariates times the degrees of freedom of the base procedure is not appropriate for assigning degrees of freedom).

We now present some simulated examples where we can evaluate the true $\mathrm{df}_{\mathrm{true}}$ for $L_2$Boosting. The first two are with componentwise linear least squares for fitting a linear model and the third is with componentwise smoothing splines for fitting an additive model. The models are

$$Y_i = \sum_{j=1}^{p} \beta_j x_i^{(j)} + \varepsilon_i, \quad \varepsilon_i \sim \mathcal{N}(0,1) \text{ i.i.d.,}$$
$$i=1,\ldots,n,$$

with fixed design from $\mathcal{N}_p(0,\Sigma), \Sigma_{i,j} = 0.5^{|i-j|}$, peff nonzero regression coefficients $\beta_j$ and with parameters

$$p = 10, \quad \mathrm{peff} = 1, \tag{1}$$

*Peter Bühlmann is Professor, Seminar für Statistik, ETH Zürich, CH-8092 Zürich, Switzerland e-mail: buhlmann@stat.math.ethz.ch. Torsten Hothorn is Professor, Institut für Statistik, Ludwig-Maximilians-Universität München, Ludwigstraße 33, D-80539 München, Germany e-mail: Torsten.Hothorn@R-project.org. Torsten Hothorn wrote this paper while he was a lecturer at the Universität Erlangen-Nürnberg.*

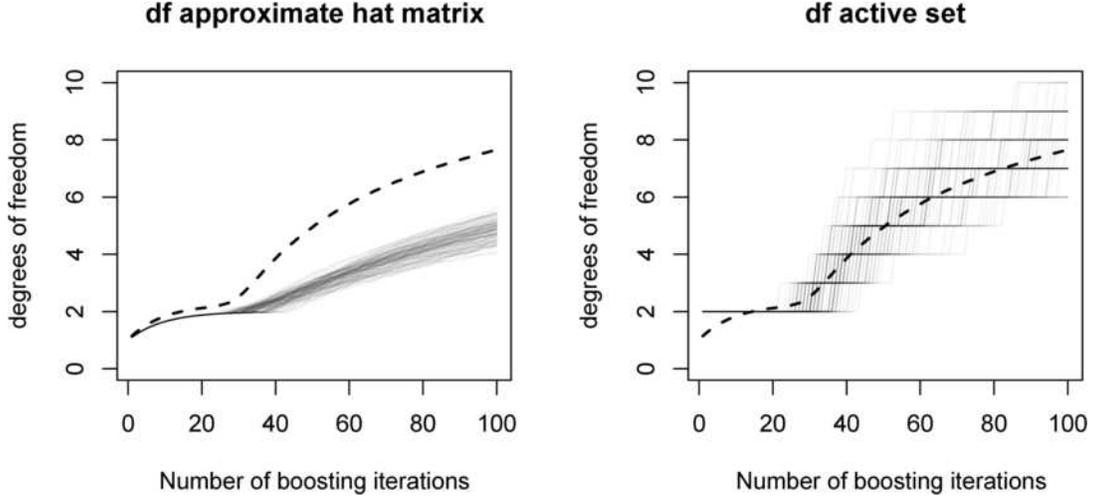

FIG. 1. *Model* (1) *and boosting with componentwise linear least squares* ($\nu = 0.1$). *True degrees of freedom* $\mathrm{df}_{\mathrm{true}}(m)$ *(dashed black line) and* $\mathrm{df}(m)$ *(shaded gray lines, left panel) and* $\mathrm{df}_{\mathrm{actset}}(m)$ *(shaded gray lines, right panel) from 100 simulations.*

$$
\begin{aligned}
&n = 100, \quad \beta_5 = \sqrt{34.5}, \quad \beta_j \equiv 0 \ (j \neq 5), \\
&p = 200, \quad \mathrm{peff} = 1, \\
(2) \quad &n = 100, \quad \beta \text{ as in } (1), \\
&p = 200, \quad \mathrm{peff} = 10, \\
(3) \quad &n = 100, \\
&\beta = (1, 1, 1, 1, 1, 0.5, 0.5, 0.5, 0.5, 0.5, 0, 0, \ldots).
\end{aligned}
$$

All models (1)–(3) have the same signal-to-noise ratio. In addition, we consider the Friedman #1 additive model with $p = 20$ and $\mathrm{peff} = 5$:

$$Y_i = 10\sin(\pi x_i^{(1)} x_i^{(2)}) + 20(x_i^{(3)} - 0.5)^2$$

$$+ 10 x_i^{(4)} + 5 x_i^{(5)} + \varepsilon_i, \quad i = 1, \ldots, n,$$

with fixed design from $\mathcal{U}[0,1]^{20}$ and i.i.d. errors $\varepsilon_i \sim \mathcal{N}(0, \sigma_\varepsilon^2)$, $i = 1, \ldots, n$ where

$$(4) \qquad \sigma_\varepsilon^2 = 1,$$

$$(5) \qquad \sigma_\varepsilon^2 = 10.$$

Figures 1–4 display the results. As already mentioned, our approximation $\mathrm{df}(m)$ underestimates the true degrees of freedom. Hence, our penalty term in AIC or similar information criteria tends to be too small. Furthermore, our $\mathrm{df}(m)$ is less variable than $\mathrm{df}_{\mathrm{actset}}(m)$. When looking in detail to the sparse cases from model (1) and (2) in Figures 1 and 2, respectively: (i) our $\mathrm{df}(m)$ is accurate for the range of

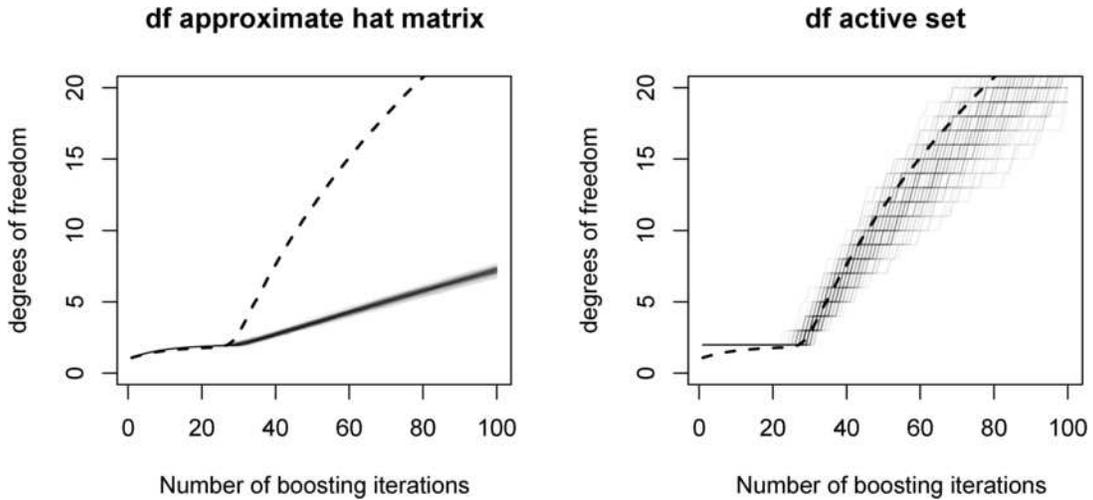

FIG. 2. *Model* (2) *and boosting with componentwise linear least squares* ($\nu = 0.1$). *Other specifications as in Figure* 1.

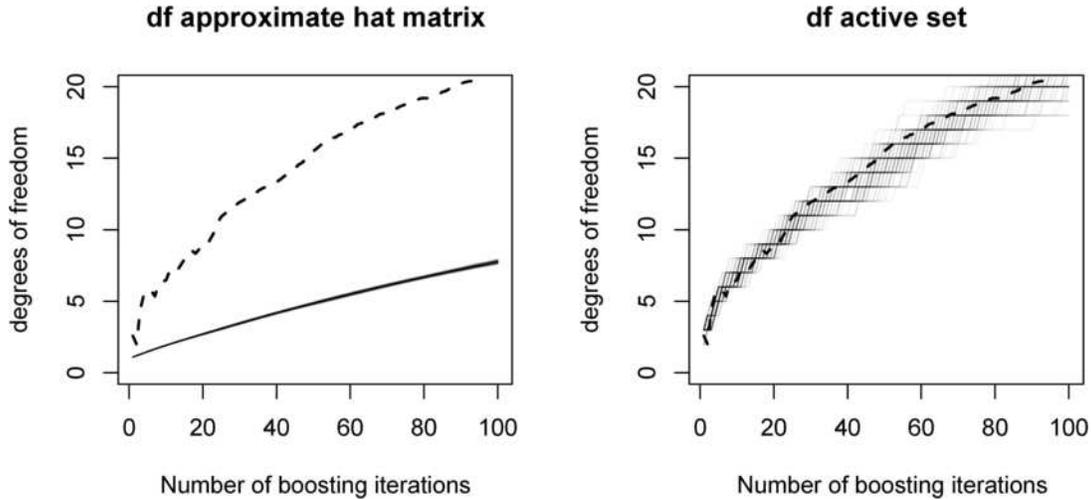

FIG. 3. *Model* (3) *and boosting with componentwise linear least squares* ($\nu = 0.1$). *Other specifications as in Figure* 1.

iterations which are reasonable (note that we should not spend more degrees of freedom than, say, 2–3 if peff = 1; OLS on the single effective variable, including an intercept, would have $\text{df}_{\text{true}} = 2$); (ii) the active set degrees of freedom are too large for the first few values of $m$, that is, $\text{df}_{\text{actset}}(m) = 2$ (one variable and the intercept) although $\text{df}_{\text{true}}(m) < 1.5$ for $m \leq 5$. Such a behavior disappears in the less sparse case in model (3), which is an example where $\text{df}(m)$ underestimates very heavily; see Figure 3.

Despite some (obvious) drawbacks of $\text{df}_{\text{actset}}(m)$, it works reasonably well. Hastie has asked us to give a correction formula for our $\text{df}(m)$. His discussion summarizing the nice relation between LARS, iF-SLR and $L_2$Boosting, together with our simulated examples, suggests that $\text{df}_{\text{actset}}(m)$ is a better approximation for degrees of freedom for boosting with the componentwise linear base procedure. We have implemented $\text{df}_{\text{actset}}(m)$ in version 1.0-0 of the **mboost** package [9] for assigning degrees of freedom of boosting with componentwise linear least squares for generalized linear models. Unfortunately, in contrast to LARS, $\text{df}_{\text{actset}}(m)$ will never be exact. It seems that assigning correct degrees of freedom for boosting is more difficult than for LARS. For other learners, for example, the componentwise smoothing spline, we do not even have a better approximation for degrees of freedom. Our formula $\text{df}(m)$ worked reasonably well for the models in (4) and (5); changing the signal-to-noise ratio by a factor 10 gave almost identical results (which is unclear a priori because $\mathrm{df}(m)$ depends on the data). But this is no guarantee for generalizing to other settings. In absence of a better approximation formula in general, we still think that our $\mathrm{df}(m)$ formula is useful as a rough approximation for degrees of freedom of boosting with componentwise smoothing splines. And we agree with Hastie that cross-validation is a valuable alternative for the task of estimating the stopping point of boosting iterations.

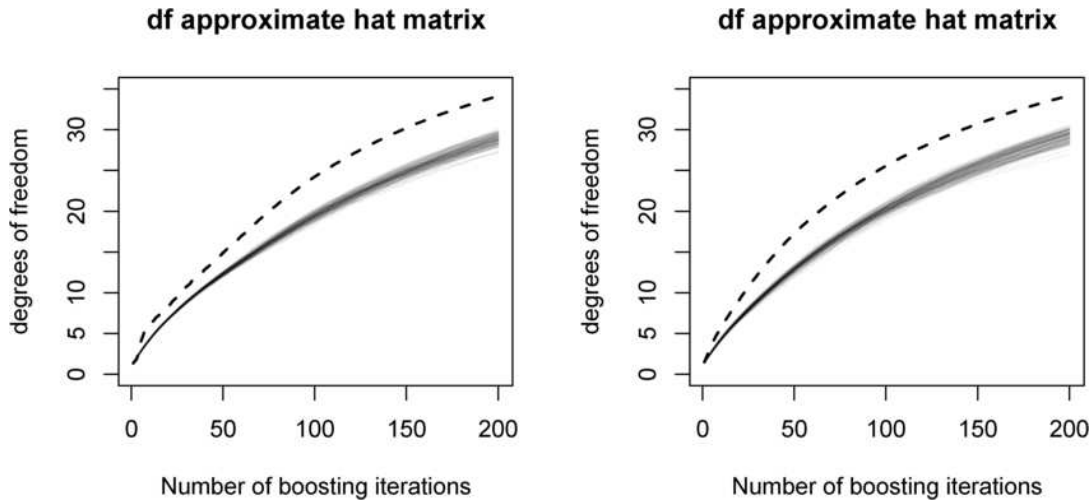

FIG. 4. *Left: model* (4). *Right: model* (5). *Boosting with componentwise smoothing splines with four degrees of freedom* ($\nu = 0.1$). *True degrees of freedom* $\text{df}_{\text{true}}(m)$ *(dashed black line) and* $\text{df}(m)$ *(shaded gray lines, for both panels)*.

## 2. HISTORICAL REMARKS AND NUMERICAL OPTIMIZATION

Buja, Mease and Wyner (BMW hereafter) make a very nice and detailed contribution regarding the history and development of boosting.

BMW also ask why we advocate Friedman's gradient descent as the boosting standard. First, we would like to point out that computational efficiency in boosting does not necessarily yield better statistical performance. For example, a small step-size may be beneficial in comparison to step-size $\nu = 1$, say. Related to this fact, the quadratic approximation of the loss function as described by BMW may not be better than the linear approximation. To exemplify, take the negative log-likelihood loss function in (3.1) for binary classification. When using the linear approximation, the working response (i.e., the negative gradient) is

$$z_{i,\mathrm{linapp}} = 2(y_i - p(x_i)), \quad y_i \in \{0, 1\}.$$

In contrast, when using the quadratic approximation, we end up with LogitBoost as proposed by Friedman, Hastie and Tibshirani [7]. The working response is then

$$z_{i,\mathrm{quadapp}} = \frac{1}{2} \frac{y_i - p(x_i)}{p(x_i)(1 - p(x_i))}.$$

The factor $1/2$ appears in [7] when doing the linear update but not for the working response. We see that $z_{i,\mathrm{quadapp}}$ is numerically problematic whenever $p(x_i)$ is close to 0 or 1, and [7], on pages 352–353, address this issue by thresholding the value of $z_{i,\mathrm{quadapp}}$ to an "ad hoc" upper limit. On the other hand, with the linear approximation and $z_{i,\mathrm{linapp}}$, such numerical problems do not arise. This is a reason why we generally prefer to work with the linear approximation and Friedman's gradient descent algorithm [6].

BMW also point out that there is no "random element" in boosting. In our experience, aggregation in the style of bagging is often very useful. A combination of boosting with bagging has been proposed in Bühlmann and Yu [2] and similar ideas appear in Friedman [8] and Dettling [4]. In fact, random forests [1] also involve some bootstrap sampling in addition to the random sampling of covariates in the nodes of the trees; without the bootstrap sampling, it would not work as well. We agree with BMW that quite a few methods actually benefit from additional bootstrap aggregation. Our paper, however, focuses solely on boosting as a "basic module" without (or before) random sampling and aggregation.

## 3. LIMITATIONS OF THE "STATISTICAL VIEW" OF BOOSTING

BMW point out some limitations of the "statistical view" (i.e., the gradient descent formulation) of boosting. We agree only in part with some of their arguments.

### 3.1 Conditional Class Probability Estimation

BMW point out that conditional class probabilities cannot be estimated well by either AdaBoost or LogitBoost, and later in their discussion they mention that overfitting is a severe problem. Indeed, the amount of regularization for conditional class probability estimation should be (markedly) different than for classification. For probability estimation we typically use (many) fewer iterations, that is, a less complex fit, than for classification. This fits into the picture of the rejoinder in [7] and [2], saying that the 0-1 misclassification loss in (3.2) is much more insensitive to overfitting. For accurate conditional class probability estimation, we should use the surrogate loss, for example, the negative log-likelihood loss in (3.1), for estimating (e.g., via cross-validation) a good stopping iteration. Then, conditional class probability estimates are often quite reasonable (or even very accurate), depending of course on the base procedure, the structure of the underlying problem and the signal-to-noise ratio. We agree with BMW that AdaBoost or LogitBoost overfit for conditional class probability estimation when using the wrong strategy—namely, tuning the boosting algorithm according to optimal classification. Thus, unfortunately, the goals of accurate conditional class probability estimation and good classification are in conflict with each other. This is a general fact (see rejoinder by Friedman, Hastie and Tibshirani [7]) but it seems to be especially pronounced with boosting complex data. Having said that, we agree with BMW that AIC/BIC regularization with the negative log-likelihood loss in (3.1) for binary classification will be geared toward estimating conditional probabilities, and for classification, we should use more iterations (less regularization).

### 3.2 Robustness

For classification, BMW argue that robustness in the response space is not an issue since, "binary responses have no problem of vertically outlying values." We disagree with the relevance of their argument. For logistic regression, robustification of the MLE has been studied in detail. Even though the MLE has bounded influence, the bound may be too large and for practical problems this may matter a lot. Künsch, Stefanski and Carroll [10] is a good reference which also cites earlier papers in this area. Note that with the exponential loss, the issue of too large influence is even more pronounced than with the log-likelihood loss corresponding to the MLE.

## 4. EXEMPLIFIED LIMITATIONS OF THE "STATISTICAL VIEW"

The paper by Mease and Wyner [11] presents some "contrary evidence" to the "statistical view" of boosting. We repeat some of the points made by Bühlmann and Yu [3] in the discussion of Mease and Wyner's paper.

### 4.1 Stumps Should be Used for Additive Bayes Decision Rules

The sentence in the subtitle which is put forward, discussed and criticized by BMW never appears in our paper. The main source of confusion seems to be the concept of "additivity" of a function. It should be considered on the logit-scale (for AdaBoost, LogitBoost or BinomialBoosting), since the population minimizer of AdaBoost, LogitBoost or BinomialBoosting is half of the log-odds ratio. Mease and Wyner [11] created a simulation model which is additive as a decision function but nonadditive on the logit-scale for the conditional class probabilities; and they showed that larger trees are then better than stumps (which is actually consistent with what we write in our paper). We think that this is the main reason why Mease and Wyner [11] found "contrary evidence."

We illustrate in Figure 5 that our heuristics to prefer stumps over larger trees is useful if the underlying model is additive for the logit of the conditional class probabilities. The simulation model here is the same as in [3] which we used to address the "contrary evidence" findings in [11]; our model is inspired by Mease and Wyner [11] but we make the conditional class probabilities additive on the logit-scale:

$$\text{logit}(p(X)) = 8 \sum_{j=1}^{5}(X^{(j)} - 0.5),$$
(6)
$$Y \sim \text{Bernoulli}(p(X)),$$

and $X \sim \mathcal{U}[0,1]^{20}$ (i.e., i.i.d. $\mathcal{U}[0,1]$). This model has Bayes error rate approximately equal to 0.1 (as in

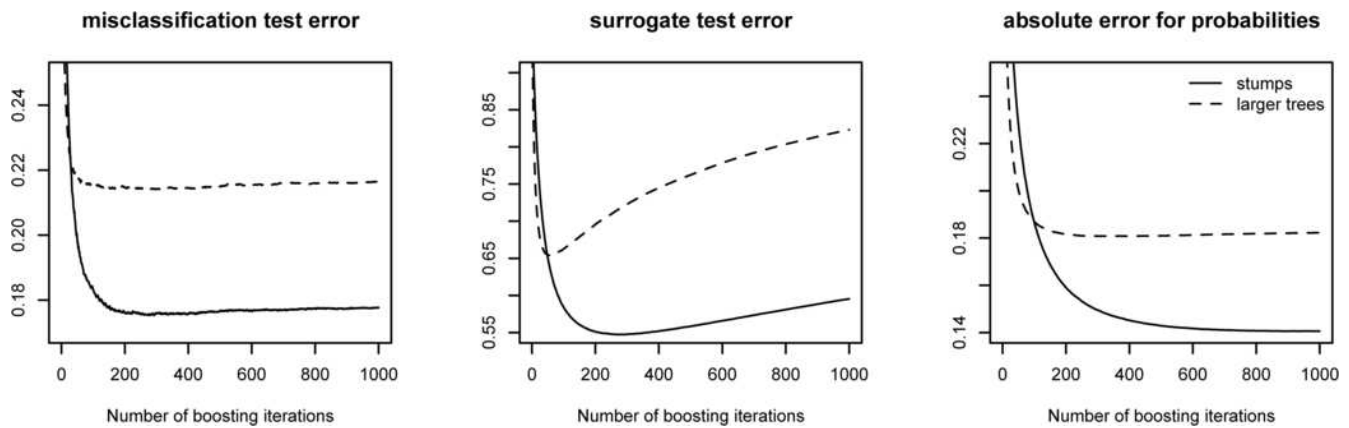

FIG. 5. *BinomialBoosting ($\nu = 0.1$) with stumps (solid line) and larger trees (dashed line) for model (6). Left panel: test-set misclassification error; middle panel: test-set surrogate loss; right panel: test-set absolute error for probabilities. Averaged over 50 simulations.*

[11]). We use $n = 100$, $p = 20$ (i.e., 15 ineffective predictors), and we generate test sets of size 2000. We consider BinomialBoosting with stumps and with larger trees whose varying size is about 6–8 terminal nodes. We consider the misclassification test error, the test-set surrogate loss with the negative log-likelihood and the absolute error for probabilities

$$\frac{1}{2000}\sum_{i=1}^{2000}|\hat{p}(X_i) - p(X_i)|,$$

where averaging is over the test set. Figure 5 displays the results (the differences between stumps and larger trees are significant) which are in line with the explanations and heuristics in our paper but very different from what BMW describe. To reiterate, we think that the reason for the "contrary evidence" in Mease and Wyner [11] comes from the fact that their model is not additive on the logit-scale. We also see from Figure 5 that early stopping is important for probability estimation, in particular when measuring in terms of test-set surrogate loss; a bit surprisingly, BinomialBoosting with stumps does not overfit within the first 1000 iterations in terms of absolute errors for conditional class probabilities (this is probably due to the low Bayes error rate of the model; eventually, we will see overfitting here as well). Finally, Bühlmann and Yu [3] also argue that the findings here also appear when using "discrete AdaBoost."

In our opinion, it is exactly the "statistical view" which helps to explain the phenomenon in Figure 5. The "parameterization" with stumps is only "efficient" if the model for the logit of the conditional class probabilities is additive; if it is nonadditive on the logit-scale, it can easily happen that larger trees are better base procedures, as found indeed by Mease and Wyner [11].

### 4.2 Early Stopping Should be Used to Prevent Overfitting

BMW indicate that early stopping is often not necessary—or even degrades performance. One should be aware that they consider the special case of binary classification with "discrete AdaBoost" and use trees as the base procedure. Arguably, this is the original proposal and application of boosting.

In our exposition, though, we not only focus on binary classification but on many other things, such as estimating class conditional probabilities, regression functions and survival functions. As BMW write, when using the surrogate loss for evaluating the performance of boosting, overfitting kicks in quite early and early stopping is often absolutely crucial. It is dangerous to present a message that early stopping might degrade performance: the examples in Mease and Wyner [11] provide marginal improvements of about 1–2% without early stopping (of course, they also stop somewhere) while the loss of not stopping early can be huge in applications other than classification.

### 4.3 Shrinkage Should be Used to Prevent Overfitting

We agree with BMW that shrinkage does not always improve performance. We never stated that shrinkage would prevent overfitting. In fact, in linear models, infinitesimal shrinkage corresponds to the Lasso (see Section 5.2.1) and clearly, the Lasso is not free of overfitting. In our view, shrinkage adds another dimension of regularization. If we do not want to tune the amount of shrinkage, the value $\nu = 0.1$ is often a surprisingly good default value. Of course, there are examples where such a default value is not optimal.

### 4.4 The Role of the Surrogate Loss Function and Conclusions From BMW

BMW's comments on the role of the surrogate loss function when using a particular algorithm are intriguing. Their algorithm can be viewed as an ensemble method; whether we should call it a boosting algorithm is debatable. And for sure, their method is not within the framework of functional gradient descent algorithms.

BMW point out that there are still some mysteries about AdaBoost. In our view, the overfitting behavior is not well understood while the issue of using stumps versus larger tree base procedures has a coherent explanation as pointed out above. There are certainly examples where overfitting occurs with AdaBoost. The (theoretical) question is whether there is a relevant class of examples where AdaBoost is not overfitting when running infinitely many iterations. We cannot answer the question with numerical examples since "infinitely many" can never be observed on a computer. The question has to be answered by rigorous mathematical arguments. For practical purposes, we advocate early stopping as a good and important recipe.

## ACKNOWLEDGMENTS

We are much obliged that the discussants provided many thoughtful, detailed and important comments. We also would like to thank Ed George for organizing the discussion.

## REFERENCES

[1] BREIMAN, L. (2001). Random forests. *Machine Learning* **45** 5–32.
[2] BÜHLMANN, P. and YU, B. (2000). Discussion of "Additive logistic regression: A statistical view," by J. Friedman, T. Hastie and R. Tibshirani. *Ann. Statist.* **28** 377–386.
[3] BÜHLMANN, P. and YU, B. (2008). Discussion of "Evidence contrary to the statistical view of boosting," by D. Mease and A. Wyner. *J. Machine Learning Research* **9** 187–194.
[4] DETTLING, M. (2004). BagBoosting for tumor classification with gene expression data. *Bioinformatics* **20** 3583–3593.
[5] EFRON, B., HASTIE, T., JOHNSTONE, I. and TIBSHIRANI, R. (2004). Least angle regression (with discussion). *Ann. Statist.* **32** 407–499. MR2060166
[6] FRIEDMAN, J. (2001). Greedy function approximation: A gradient boosting machine. *Ann. Statist.* **29** 1189–1232. MR1873328
[7] FRIEDMAN, J., HASTIE, T. and TIBSHIRANI, R. (2000). Additive logistic regression: A statistical view of boosting (with discussion). *Ann. Statist.* **28** 337–407. MR1790002
[8] FRIEDMAN, J. H. (2002). Stochastic gradient boosting. *Comput. Statist. Data Anal.* **38** 367–378. MR1884869
[9] HOTHORN, T., BÜHLMANN, P., KNEIB, T. and SCHMID, M. (2007). **Mboost**: Model-based boosting. R package version 1.0-0. Available at http://CRAN.R-project.org.
[10] KÜNSCH, H.-R., STEFANSKI, L. A. and CARROLL, R. J. (1989). Conditionally unbiased bounded-influence estimation in general regression models, with applications to generalized linear models. *J. Amer. Statist. Assoc.* **84** 460–466. MR1010334
[11] MEASE, D. and WYNER, A. (2008). Evidence contrary to the statistical view of boosting. *J. Machine Learning Research* **9** 131–156.